\documentclass[conference]{IEEEtran}
\IEEEoverridecommandlockouts
\usepackage{cite}
\usepackage{amsmath,amssymb,amsfonts}
\usepackage{algorithmic}
\usepackage{graphicx}
\usepackage{textcomp}
\usepackage{xcolor}
\usepackage{float}
\usepackage[caption = false]{subfig}

\def\BibTeX{{\rm B\kern-.05em{\sc i\kern-.025em b}\kern-.08em
    T\kern-.1667em\lower.7ex\hbox{E}\kern-.125emX}}
    
    \newcommand{\vet}[1]{\mathbf{#1}}

\newcommand{\MAHF}{Multiscale Anisotropic Harmonic Filters}

\begin{document}

\title{Multiscale Anisotropic Harmonic Filters \\ on  non Euclidean domains
}

\author{\IEEEauthorblockN{ Francesco Conti, Gaetano Scarano, Stefania Colonnese}
\IEEEauthorblockA{\textit{Dept. of Information, Electronics and Telecommunication, University of Rome "Sapienza"} }
\thanks{Submitted to EUSIPCO.
\newline
F.~Conti, ~G.~Scarano, S.~Colonnese are Department
 of Information Engineering, Electronics and Telecommunications, e-mail: conti.1655885@studenti.uniroma1.it, 
 gaetano.scarano@uniroma1.it, stefania.colonnese@uniroma1.it
.}
}

\maketitle

\begin{abstract}
This paper introduces \MAHF \:(MAHFs)
aimed at extracting signal variations over non-Euclidean domains, namely 2D-Manifolds and their discrete representations, such as meshes and 3D Point Clouds as well as graphs. The topic of pattern analysis is central in image processing and, considered the growing interest in new domains for information representation, the extension of analogous practices on volumetric data is highly demanded. To accomplish this purpose, we define MAHFs as the product of two  components, respectively related to a suitable smoothing function, namely the heat kernel derived from the heat diffusion equations, and to local directional information. 
We analyse the effectiveness of our approach in multi-scale filtering and variation extraction. Finally, we present an application to the surface normal field and to a luminance signal textured to a mesh, aiming to spot, in a separate fashion, relevant curvature changes (support variations) and signal variations.
\end{abstract}


\section{Introduction}
 Signal and image processing tools, such as for instance Fourier transform or wavelet, typically  rely, implicitly or explicitly, on a measure of Euclidean distance in signal's domain \cite{Bro, Ric}. This metric is not significant on domains involved in immersive multimedia systems to represent volumetric data, namely 2D-manifolds and their discrete representations, as meshes and 3D-Point Clouds (3D-PCs). On the other hand, Signals (defined) on Graphs (SoGs) provide a powerful and effective model for data defined on the aforementioned supports \cite{Ortega}. Indeed, Graph Signal Processing (GSP) has found application  in all the branches of signal processing on volumetric domains, for the purpose of dual description and filtering \cite{Tan}, enhancement and restoration \cite{Dinesh}, compression \cite{Thanou}, and transmission \cite{Fuji}.

In this paper, we present  a novel filtering approach to extract signals variations on non-euclidean domains. Specifically, we present a class of tools which we denote as  \MAHF\:(MAHFs), aimed at identifying specific patterns (edges, forks) on textures manifolds. 
In image processing, specific angular patterns  are detected by Circular Harmonic Filters (CHFs) \cite{Jacovitti},  used for example in deconvolution \cite{Panci} or interpolation \cite{CHF_interp}, due to their  invariance and spectral properties. These properties are achieved exploiting relevant information characterizing two pixels: $i)$ their distance, that allows to define smoothing kernels in the radial variable, $ii)$ the direction of the joining segment. The radial function and the angular information are spent in a separable fashion to construct CHFs.
In MAHFs,  radial and angular information are tuned non-Euclidean geometries. As far as  the smoothing function is concerned, we resort to the concept of heat kernel, derived from the heat diffusion equations. On the other hand, angles are measured using projections on the tangent plane, as recently carried on in anisotropic or rotation invariant CNNs on meshes \cite{Wiersma, de_Haan}.

The herein defined filters are suited to be employed in  content analysis,  as salient points detection, shape classification. Furthermore, the extraction of specific  patterns can be leveraged as a pre-processing stage for signal enhancement as well as adaptive encoding.

To summarize, the main contributions here presented are:
$i)$ the introduction of a novel approach for anisotropic filtering on volumetric data, leveraging angular measures recently introduced in geometry processing and Deep Learning on manifolds; $ii)$ the design of MAHFs, novel multi-scale filters capable to extract signal variations on 3D-shapes;  $iii)$ the application of MAHFs to spot local variations of the normal field and luminance signals separately.

The organization of the paper is as follows: in Sec. \ref{sec:signal_processing} we recall the basic elements to process signals on non-Euclidean domains; in Sec.\ref{sec:MHF} we introduce MAHFs; in Sec.\ref{sec:volume} we experimentally verify the effectiveness of the method presenting several applications on signals defined over real volume objects; Sec.\ref{sec:concl} concludes the paper.

\section{Signal Processing on Non-Euclidean Domains}
\label{sec:signal_processing}
Let us consider a 2D manifold $\mathcal{M}$  embedded in a 3D domain $\mathbb{R}^3$, usually referred as 3D-shape, and a signal on the manifold $u(\textbf{p}), \textbf{p}\!=\!(x,y,z)\!\in\!\mathcal{M}$, which can represent information as intensities or directions. Classical signal processing tools can be re-defined on manifolds under suitable smoothness hypothesis. In particular, considering the set of eigenfunctions $\chi_s: \mathcal{M} \rightarrow \mathbb{R}$ and eigenvalues $\lambda_s$ of the Manifold's Laplace-Beltrami Operator (LBO) $\Delta$, it is possible to define a basis for a Fourier-like transform, the so called Manifold Harmonics \cite{Vallet, Bro}. The same procedure, by discretizing the LBO, can be accomplished for signals defined over discrete 3D-shapes, namely 3D-PCs and meshes. The latter ones can be considered piece-wise approximations of a manifold, composed by triplets $(\mathcal{V},\mathcal{E}, \mathcal{F})$, where $\mathcal{V} \!\subset \!\mathbb{R}^{3 \times N}$ collects the vertices set, also referred as 3D-PC, while $\mathcal{F}\! \subset\! \mathcal{V}\! \times\! \mathcal{V}\! \times \!\mathcal{V}$ (i.e. triangles) and $\mathcal{E}\! \subset\! \mathcal{V} \times \!\mathcal{V}$ collect faces and edges. 

GSP  provides an effective framework to process signal defined over manifolds' discrete representation \cite{Tan}. In GSP, the signal domain is a graph, defined as  $\mathcal{G}\!=\!(\mathcal{V},\mathcal{E})$, where $\mathcal{V}$ denotes the set of vertices $v_i, i = 1, \dots, N$, and $\mathcal{E}$ the set of edges $e_{ij}$, associated with a real weight $w_{ij}$. 
Let us denote as $\mathbf{W}\!=\!\{w_{ij}\}_{i,j=1}^N$  the $N\times N$ real matrix collecting all the edge weights, known as the graph adjacency matrix. The Graph Laplacian (GL) is defined as $\mathbf{L}\!=\!\mathbf{D} - \mathbf{W}$,\footnote{Alternative definitions use different normalizations and symmetry \cite{Shuman}} where $\mathbf{D}$ is the diagonal degree matrix, having $d_{ii}\!=\!\sum_{j=1}^{N}w_{ij}$, for all $i\!=\!1,\ldots, N$. The GL's eigenvectors form a basis for a Fourier-like transform \cite{Shuman}. 

Given a mesh or a 3D-PC, the GL obtained using weights properly defined according to geometrical and/or topological information corresponds to a LBO discretization. As far 3D-PCs are considered \cite{Dinesh, Liang}, weights are commonly defined with Gaussian functions of the Euclidean distance between k-Nearest Neighbor points, exploiting only the geometrical information. Differently, concerning triangular meshes \cite{Tan}, the weights $w_{ij}$ associated to the edge $e_{ij}$ connecting $i$ and $j$ are defined leveraging also topology as $w_{ij}\!=\! \left(\cot{\alpha_{ij}} + \cot{\beta_{ij}} \right)/2\mathcal{A}_i$,
where $\alpha_{ij}$ and $\beta_{ij}$ denote the angles opposite to $e_{i, j}$, while $\mathcal{A}_i$ is the estimated area of the triangles adjacent to the vertex $i$\cite{Huang_2020}.
The above elements provide  the ground to extend signal processing tools in all the non-Euclidean domains. A relevant example of this extension is the definition of the Spectral Graph Wavelet Transform \cite{Hammond} on volumetric data \cite{Tan}.

\section{\MAHF}
\label{sec:MHF}
Herein, we introduce our novel \MAHF \:aiming to extract variation of signals defined over continuous and discrete manifolds.
Anisotropic filtering and signal variations detection have been largely researched in image processing and, as a significant example, we refer to CHFs \cite{Panci}, \cite{CHF_interp}. Consider an image, modeled on a continuous domain by a real function $f(x,y)$; using the polar representation, a CHF of order $k$ is a complex filter defined as \cite{Panci}:
\vspace*{-0.3em}
\begin{equation}\label{eq:chf}
    \psi^{(k)}(r,\theta) = w_k(r) e^{jk\theta }
\end{equation}
where the dependence from the radial and the angular components are separated. The functions $w_k(r)$ are typically chosen as isotropic Gaussian kernels in the variable $r$, in order to guarantee isomorphism with the frequency domain.
For increasing $k$, CHFs are known to highlight image structures of increasing complexity, as edges ($k\!=\!1$), corners ($k\!=\!2$), forks ($k\!=\!3$) and crosses ($k\!=\!4$) \cite{Panci}. In particular, filter output's module is expressive of structure intensity, while the phase contains information about directionality. Further, a scaling factor $\alpha$ allows to define a multi-resolution analysis, composing a Circular Harmonic Wavelets (CHWs) basis \cite{Jacovitti}. 

Inspired by CHFs, in order to introduce our novel approach, we need to establish: 1) a smoothing kernel replicating the isotropic gaussian smoothing on a 2D-manifold; 2) an angular measurament $\theta$ on the manifold surface.

\subsection{MAHFs on Manifolds}\label{subsec:MAHF_manifolds}
We firstly introduce MAHFs on a continuous shape $\mathcal{M}$ with LBO $\Delta$. As concern the smoothing kernel, we refer to the heat kernel \cite{Bro, Kirgo}, the fundamental solution of the heat equation:
$$    \Delta u(\textbf{p}, \: t)\! = \!- \partial_t u(\textbf{p}, \: t), \quad  u(\textbf{p},\: t\!=\!0) \!=\! u_0(\textbf{p})
$$
where $u(\textbf{p}, \: t), \: \textbf{p}\!\in\! \mathcal{M}, \: t\!\in\! \mathbb{R}^+$ denotes the solution starting from an initial condition $u_0(\textbf{p})$. When the initial condition consists in an impulsive function $u_0(\textbf{p})\! =\! \delta(\textbf{p}\!-\! \textbf{p}_0)$ concentrated in $\textbf{p}_0$, the solution is therefore indicated as heat kernel and denoted as $K_t(\textbf{p}, \textbf{p}_0)$, describing the heat diffused from $\textbf{p}_0$ to $\textbf{p}$ after $t(sec)$. 
Referring to the eigenfunctions
$\chi_s:\! \mathcal{M}\! \rightarrow\! \mathbb{R}$ and eigenvalues $\lambda_s$, the heat kernel can be expressed as:
\vspace*{-0.3em}
\begin{equation}\label{eq:heat_kernel}
    K_t(\textbf{p}, \textbf{p}_0) \!=\! \sum_{s=0}^{\infty}e^{-t\lambda_s} \chi_s(\textbf{p})\chi_s(\textbf{p}_0)
\end{equation}
As far a general $u_0(\textbf{p})$ is considered, we can express the solution of \eqref{eq:solution_heat} as the convolution in the eigenfunction domain with the heat kernel \cite{Huang_2020}:
\vspace*{-0.3em}
\begin{equation}\label{eq:solution_heat}
    u(\textbf{p}, \: t) \!=\! \left(K_t \ast u_0 \right)(\textbf{p})\!=\! \sum_{s=0}^{\infty}e^{-t\lambda_s} \hat{u}_0(s)\chi_s(\textbf{p})
\end{equation}
where $\hat{u}_0$ is the representation of $u_0$ in the manifold eigenfunction basis.
\begin{figure}[t]
\centering
\subfloat[$t\!=\!5 s$]
{\includegraphics[trim=3cm 0cm 3cm 2cm,clip=true,width=0.3\linewidth]{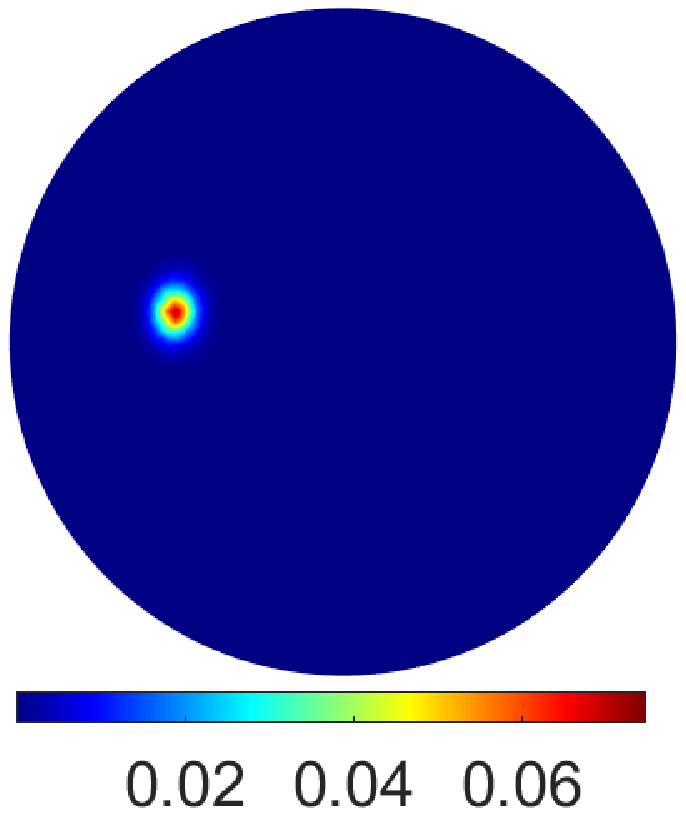}} \quad
\subfloat[$t\!=\!25 s$]
{\includegraphics[trim=3cm 0cm 3cm 2cm,clip=true,width=0.3\linewidth]{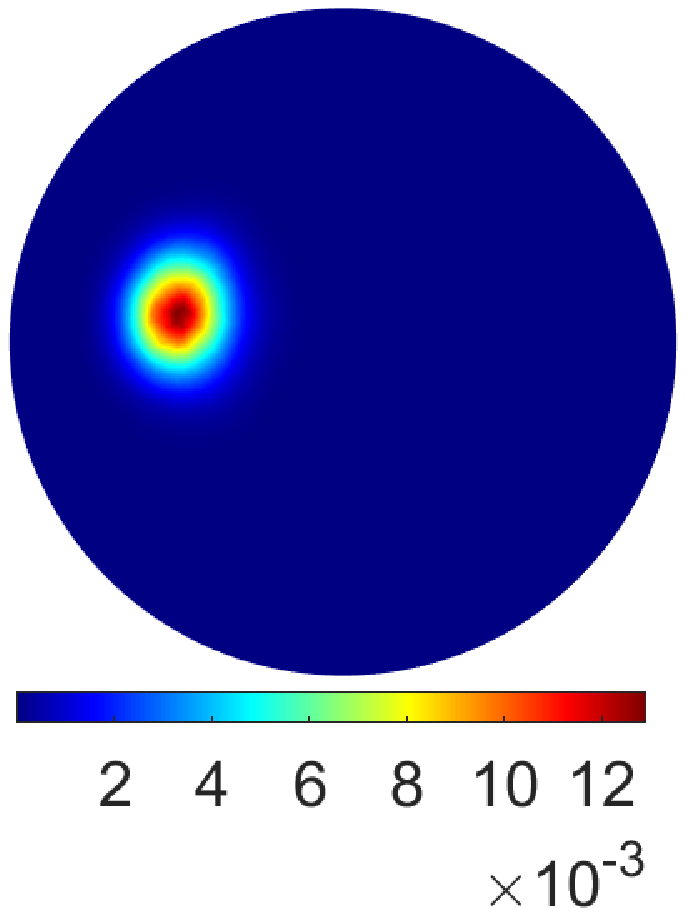}} \\
\subfloat[$t\!=\!50 s$]
{\includegraphics[trim=3cm 0cm 3cm 2cm,clip=true,width=0.3\linewidth]{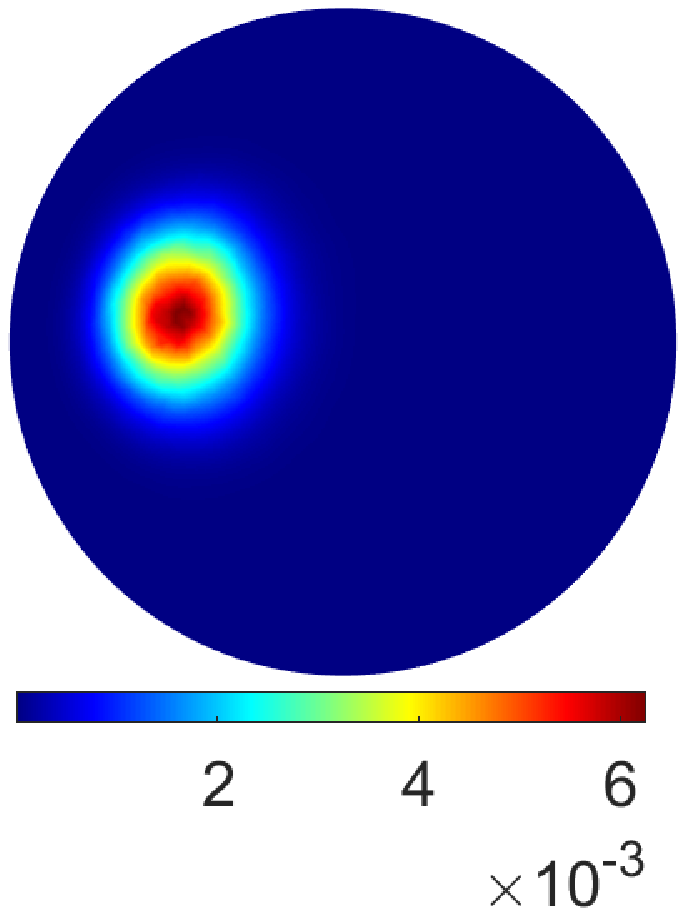}} \quad
\subfloat[$t\!=\!100 s$]
{\includegraphics[trim=3cm 0cm 3cm 2cm,clip=true,width=0.3\linewidth]{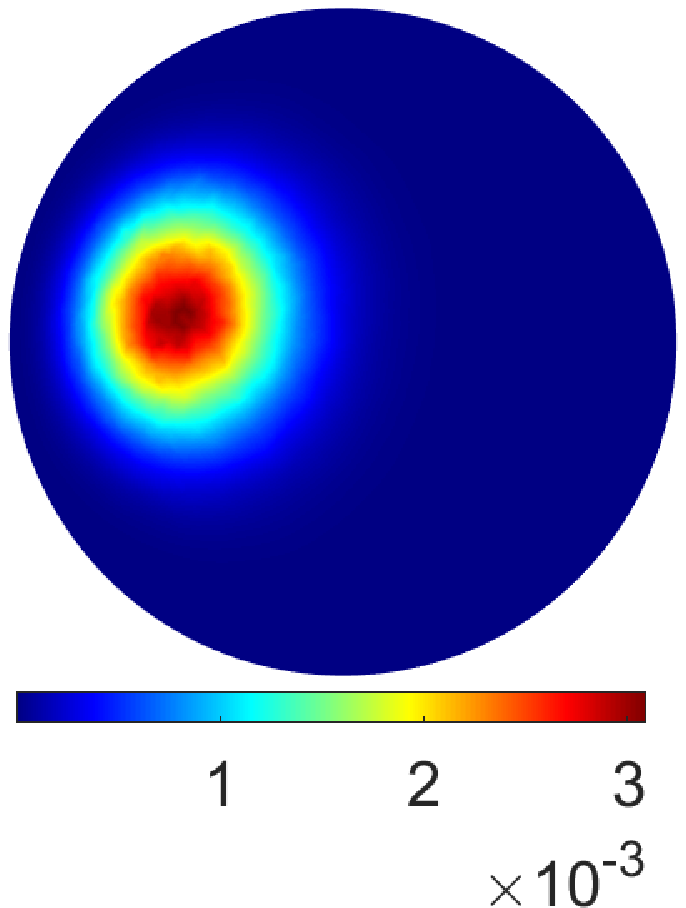}}
\caption{Heat Kernel on a sphere, decaying as a Gaussian  function as much smooth as $t$ increases}
\label{fig:heat_kernel}
\end{figure}
The heat kernel is known to define an isotropic smoothing function on $\mathcal{M}$, decaying as a Gaussian function and more smoothly with increasing $t$ \cite{Tingbo}, as can be appreciated in Fig.\ref{fig:heat_kernel} concerning a spherical 3D-shape, paving the ways to multi-scale analysis. Additionally from \eqref{eq:heat_kernel}-\eqref{eq:solution_heat}, we can deduce the Fourier representation $\hat{K}_t(s)\: =\: e^{-t\lambda_s}$, namely a Gaussian function in the frequency domain expressed wrt the frequency-like variable $\sqrt{\lambda_s}$, supplying the isomorphic property between the two domains. These important properties can be exploited to define diffusion wavelets \cite{Coifman_2006} on manifolds and graphs, based on the concept "\textit{to scale is equivalent to diffuse}"; moreover, the kernels' derivatives with respect to the variable $t$ are exploited in Computer Graphics communities to define the Mexican Hat Wavelets (MHWs) \cite{Kirgo, Tingbo}, inherently isotropic filters mainly used in shape matching and geometry processing. MHWs 2D-filters are widely applied in image processing for edge detection and thus will be taken as comparison in Sec. \ref{sec:volume}.

As concern angles, different definitions have been proposed in the context of CNNs generalization on manifolds in an anisotropic fashion (mainly limited to meshed domains and not 3D-PC). More in details, local geodesic polar coordinates \cite{Masci_2015} are defined in meshed domains using angular bins. On the other hand, alternative approaches use logarithmic mapping to the tangent plane \cite{de_Haan, Wiersma}, namely a representation of points on the plane $T_p\mathcal{M}$ tangent to the manifold in $\textbf{p}$, thus allowing an extension also to 3D-PCs.
Similarly to the latter paradigm, we define $\delta\theta(\textbf{p}; \textbf{p}_0)$ as the azimuth in spherical coordinates of $\textbf{p}$, measured wrt a frame system ($x'$, $y'$, $z'$) centered at the point $\textbf{p}_0$ and with the $z'$ axis directed with the surface normal $\textbf{n}_{p_0}$ at the point. Subsequently, the axes $x'$ and $y'$ are arbitrarily chosen on the tangent plane. Note that this choice has no effect in filtering, since the variations are extracted on two orthogonal directions, as discussed in the following.

At this point, inspired by \eqref{eq:chf}, we define the MAHFs of order $k$ around a point $\textbf{p}_0\in\mathcal{M}$ as a function $\mathcal{\psi}:\mathcal{M}\times\mathcal{M}\rightarrow \mathbb{R}^2$:
\vspace*{-0.3em}
\begin{equation}\begin{split}
    &\psi_R^{(k)}(\textbf{p}; \textbf{p}_0) \!=\!  
  K_t(\textbf{p}, \textbf{p}_0) \cos{\left(k\delta\theta(\textbf{p}; \textbf{p}_0)\right)}, \\
    &\psi_I^{(k)}(\textbf{p}; \textbf{p}_0)\! =\!  
  K_t(\textbf{p}, \textbf{p}_0) \sin{\left(k\delta\theta(\textbf{p}; \textbf{p}_0)\right)};
\end{split}
\label{eq:manifold_filters}
\end{equation}
\vspace*{-0.3em}

\begin{figure}[t]
    \centering
    \includegraphics[trim=0.3cm 0.1cm 0cm 0cm,clip=true,width=0.5\linewidth]{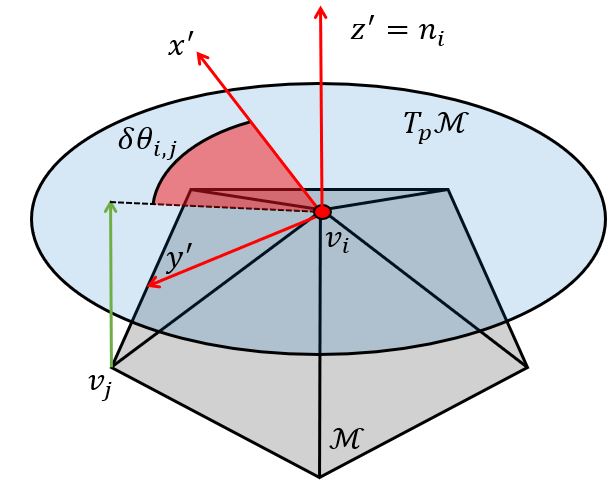}
    \caption{Angles measurement on a discrete manifold $\mathcal{M}$}
    \vspace*{-0.6em}
    \label{fig:angles}
\end{figure}
\subsection{MAHFs on graphs}\label{subsec:MAHF_graph}
We now move to MAHFs' definition for signals on discrete domains, namely meshes or 3D-PCs, represented by graphs $\mathcal{G}=(\mathcal{V},\mathcal{E})$ with the GL matrix $\textbf{L} \in \mathbb{R}^{N \times N}$ composed as exposed in Sec. \ref{sec:signal_processing}, whose $N$ eigenpairs $\varphi_s$ and $\lambda_s$ determine the Graph Fourier decomposition. In the discrete settings, we refer to the heat kernel considering the whole spectrum \cite{Bron_alex}:
\begin{equation}\label{eq:heat_kernel_discrete}
    K_t[v_i, v_j] \!= \! \sum_{s=0}^{N-1}e^{-t\lambda_s} \varphi_s[v_i]\varphi_s[v_j]
\end{equation}
The use of all the eigenpairs is computationally impractical for large $N$; for this reason, Chebyshev polynomial approximations are commonly used to evaluate \eqref{eq:heat_kernel_discrete} as a polynomial in the variable $\textbf{L}$ (see for instance \cite{Huang_2020, Patane}). It is worth to note that, referred to the multi-scale approach, the heat kernel for different $t$ values can be conveniently evaluated in a recursive manner $ K_{t_1+t_2}\! = \! K_{t_1} \ast K_{t_2}$ \cite{Huang_2020}. Additionally, \cite{Kirgo} relates the parameter $t$ to the global area of the discrete manifold $\mathcal{M}$ in multi-scale analysis.

As concern angles, also in the discrete setting we refer to the azimuth angle measured on the tangent plane to the vertex $v_i$. More in detail, we arbitrary choose 2 directions $(x', \:y')$ orthogonal to the normal $\textbf{n}_i$, \footnote{We underline that normal vectors could derive from scanning or, alternatively, could be estimated both for meshes and 3D-PCs.} as shown in Fig.\ref{fig:angles}.
\begin{figure}[t]
\centering
\subfloat[$k = 0$]
{\includegraphics[trim=1.5cm 1cm 1.2cm 2cm,clip=true,width=0.47\linewidth]{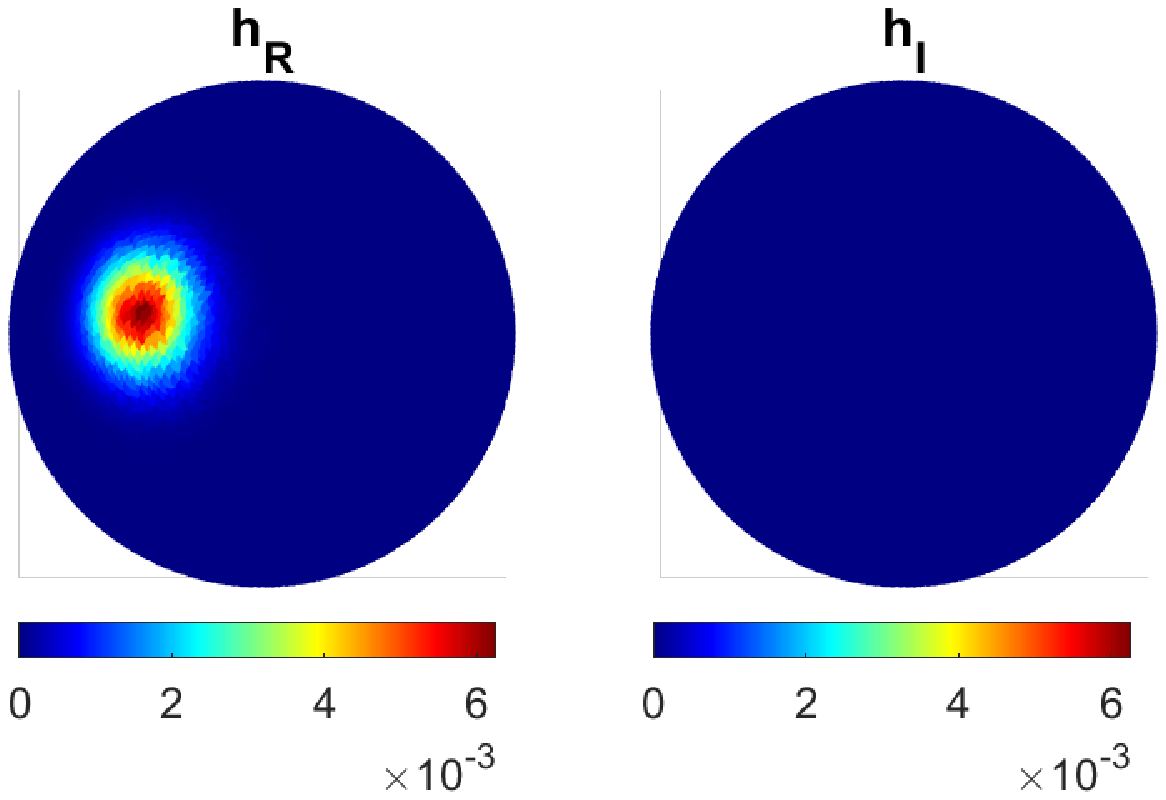}} \quad
\subfloat[$k = 1$]
{\includegraphics[trim=1.5cm 1cm 1.2cm 2cm,clip=true,width=0.47\linewidth]{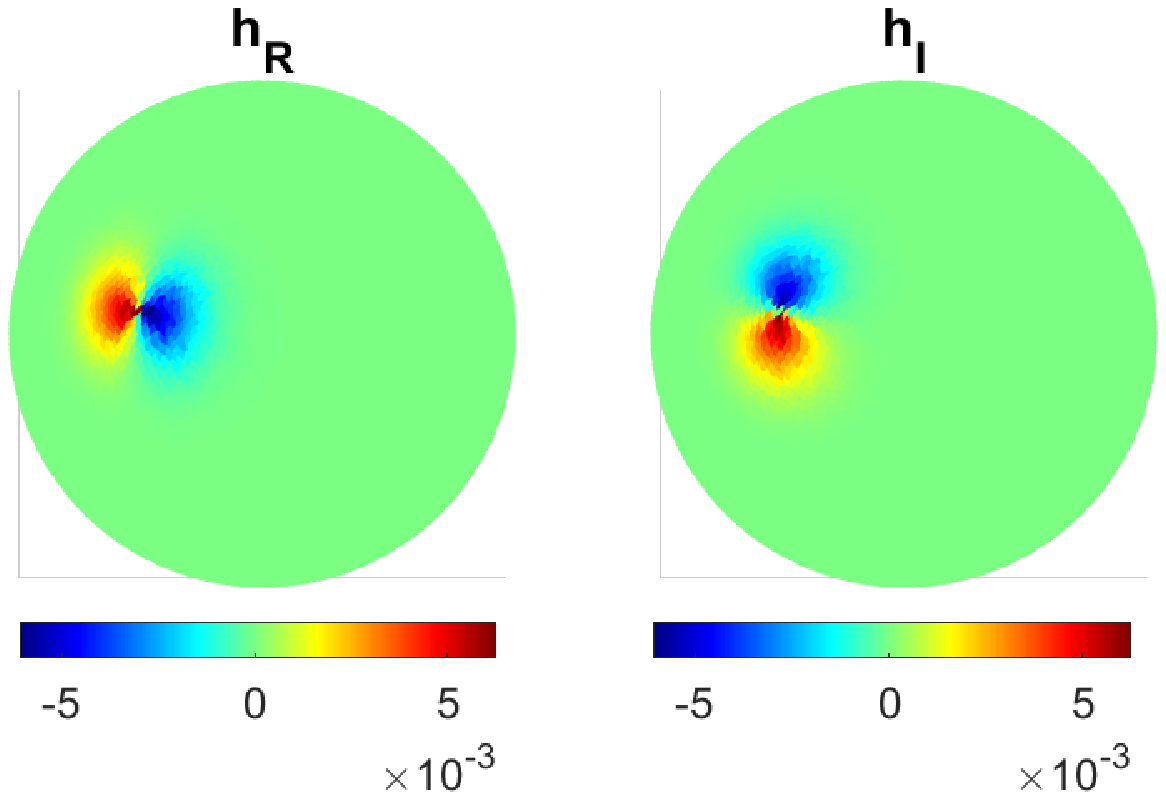}} \\
\subfloat[$k = 2$]
{\includegraphics[trim=1.5cm 1cm 1.2cm 2cm,clip=true,width=0.45\linewidth]{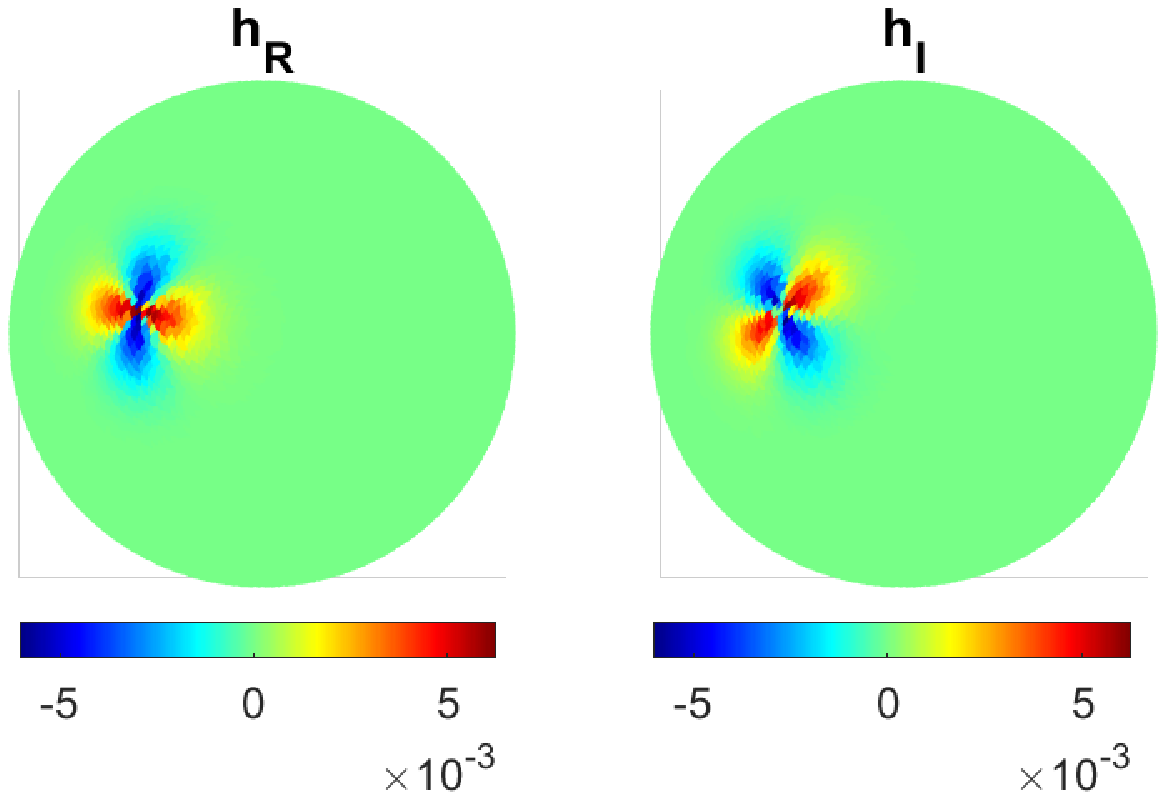}} \quad
\subfloat[$k = 3$]
{\includegraphics[trim=1.5cm 1cm 1.2cm 2cm,clip=true,width=0.45\linewidth]{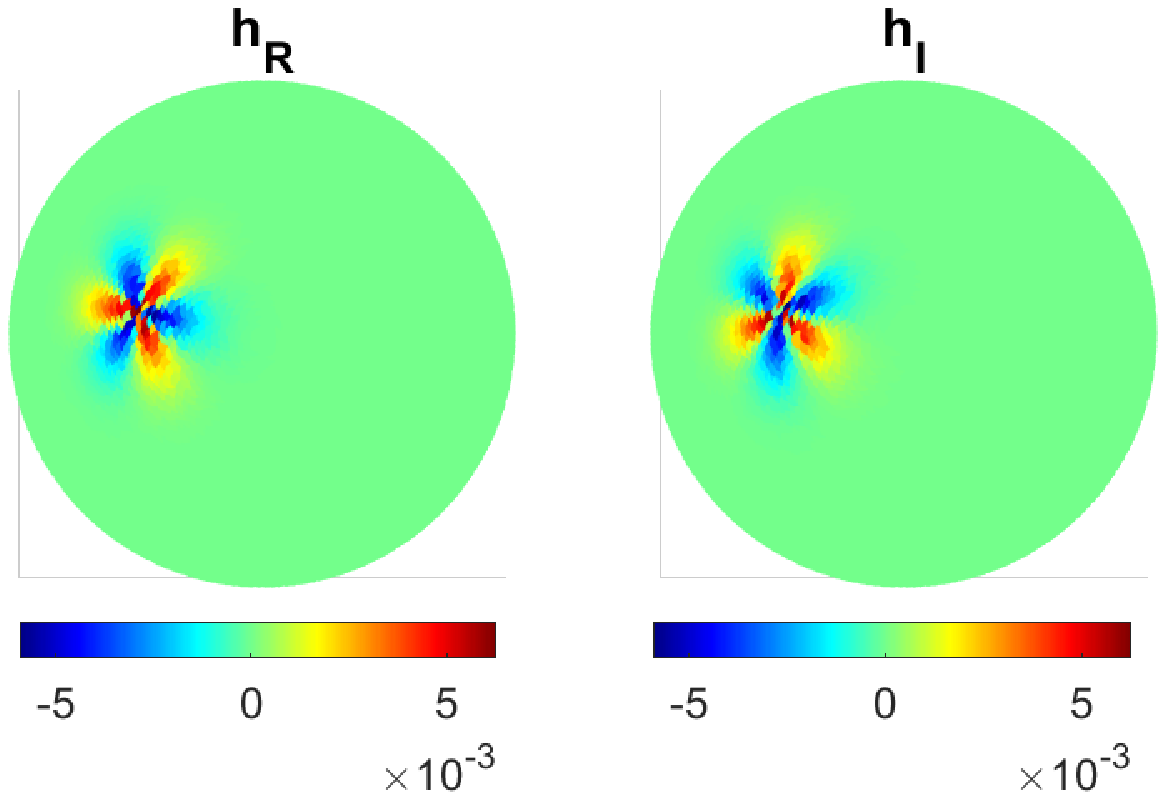}} 
\caption{MAHFs on a sphere, $t\!=\!50s$ for increasing $k$; the anisotropic behavior resembles CHFs' action on images }
\vspace*{-0.6em}
\label{fig:filters_sphere}
\end{figure}

Finally, the real and imaginary part of MAHFs on a graph
\begin{equation}\begin{split}
    &h_R^{(k)}[i, j] =K_t[v_i, v_j] \cos(k\delta\theta_{i,j}),  \\
    &h_I^{(k)}[i, j] = K_t[v_i, v_j] \sin(k\delta\theta_{i,j})
\end{split}
\label{eq:hxyz}
\end{equation}
When applied to signals defined over discrete manifolds, MAHFs provide a differential computation to extract features with higher complexity as $k$ increases, resembling the typical behavior of CHFs on images. This is illustrated in Fig.\ref{fig:filters_sphere}, where  MAHFs in \eqref{eq:hxyz} are reported for different values of $k$ on a sphere with a fixed $t = 50s$. 

Starting from the classical definition of graph vertex filtering, we now introduce the application of MAHF to signals $s[i],\;i=0,\dots N-1$ defined over graphs vertices $v_i,\;i=0,\dots N-1$. Let us denote by $r_R[i], r_I[i] \;i=0,\dots N-1$ the output signal, which is computed as follows:
\begin{equation}
 \begin{split}
     r_R[i]=\sum\limits_{j} h_R^{(k)}[v_i, v_j] s[j],  \quad
     r_I[i]=\sum_j  h_I^{(k)}[v_i, v_j] s[j]
 \end{split}
\end{equation}
We further underline that $K_t[v_i, v_j]$ defines a patch on the graph, so that the filtering in the vertex domain, though formally extended over the whole graph, can be efficiently accomplished by using only the localized support determined by thresholding the kernel.
As described for images, the square module of the output components $R^2[i]\!= \!r_R^2[i]\!+\!r_I^2[i]$
is able to capture local variability behaviors of the signal on its support. In addition, we remark that consistent angular information could be recovered using the local reference frame (previously arbitrarily selected).

\section{Application of MHF to volumetric data}
\label{sec:volume}
Herein, we presents three MAHFs applications aiming to demonstrate: $i)$ signal variations detection and multi-scale behavior; $ii)$ Normal field's local variation estimate, providing a comparisons with MHWs \cite{Kirgo}; $iii)$ edge extraction for an image textured on a face shape.

\begin{figure}[t]
\centering
\subfloat[Input signal \label{fig:signal_a}]{
\includegraphics[trim=4cm 1cm 1cm 1.5cm,clip=true,width=0.35\linewidth]{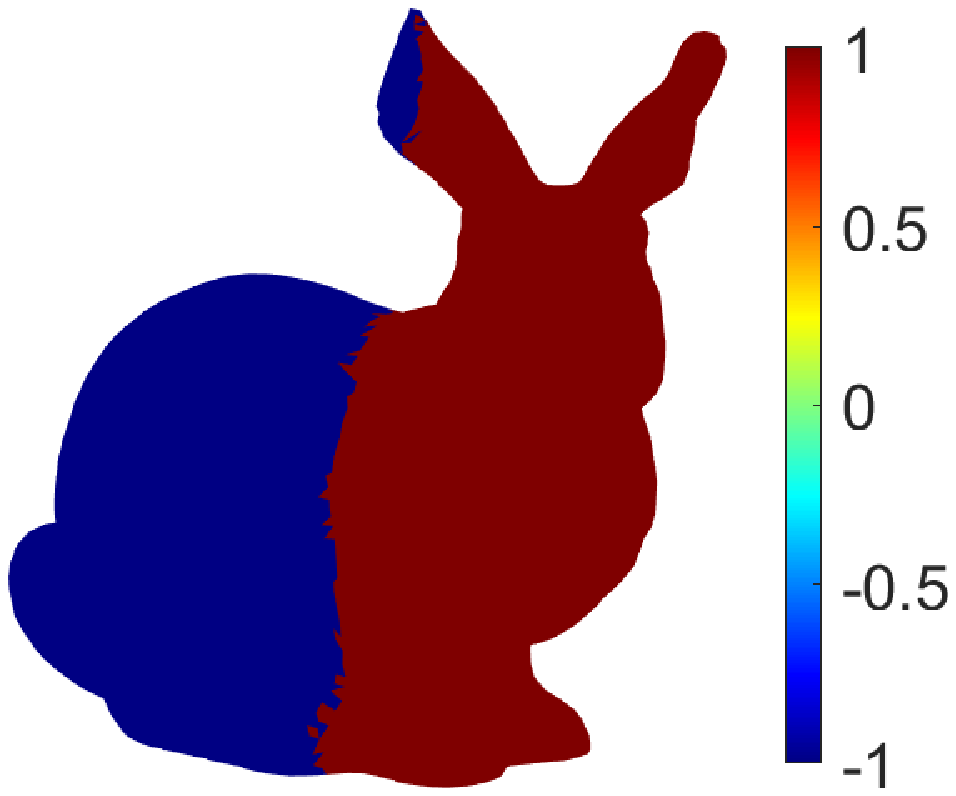}} \quad
\subfloat[Output Signal ${R^2[i]}$ \label{fig:signal_b}]{
\includegraphics[trim=4cm 1cm 1cm 1.5cm,clip=true,width=0.35\linewidth]{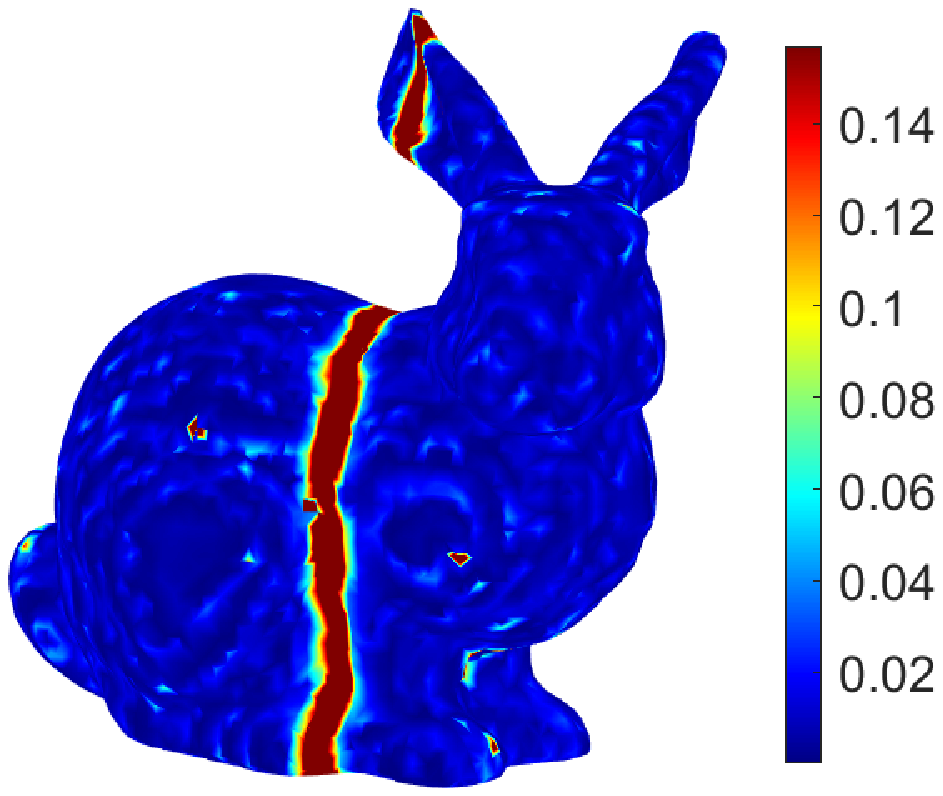}} 
\caption{MAHFs applied on a 2 levels signal}
\label{fig:signal}
\end{figure}

\begin{figure}[t]
\centering
\subfloat[Input Signal \label{fig:multiscale_signal}]
{\includegraphics[trim=4cm 1.3cm 0.9cm 1.6cm,clip=true,width=0.35\linewidth]{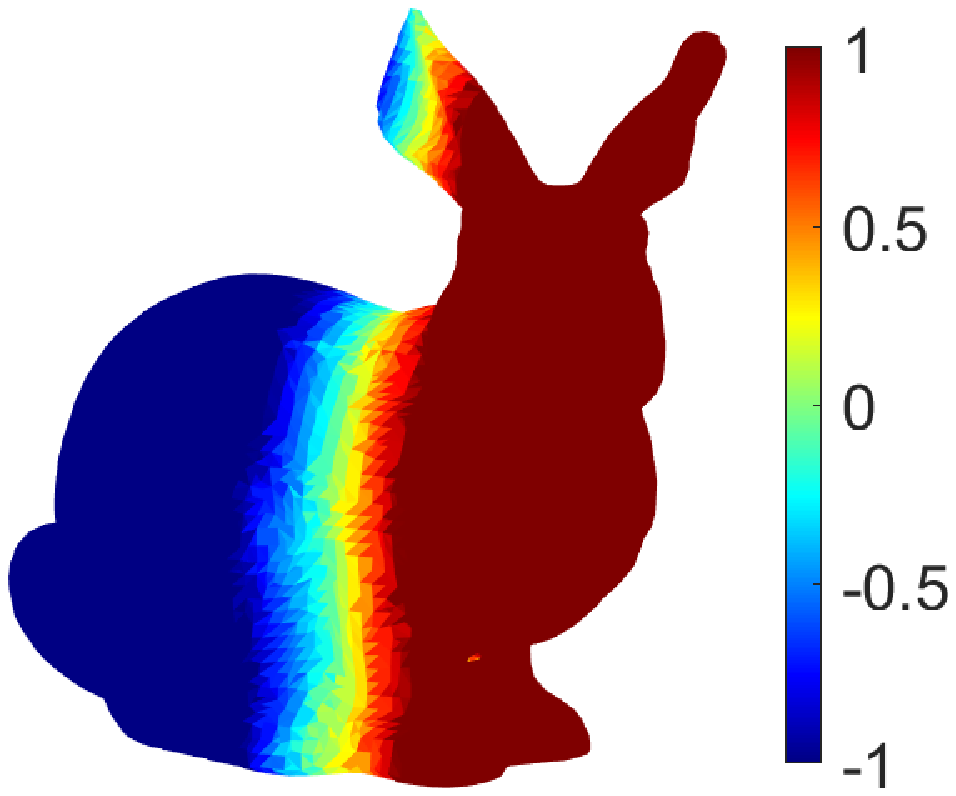}} \\
\subfloat[$t\!=\!5 s$ \label{fig:multiscale_t5}]
{\includegraphics[trim=4cm 1.3cm 0.9cm 1.6cm,clip=true,width=0.35\linewidth]{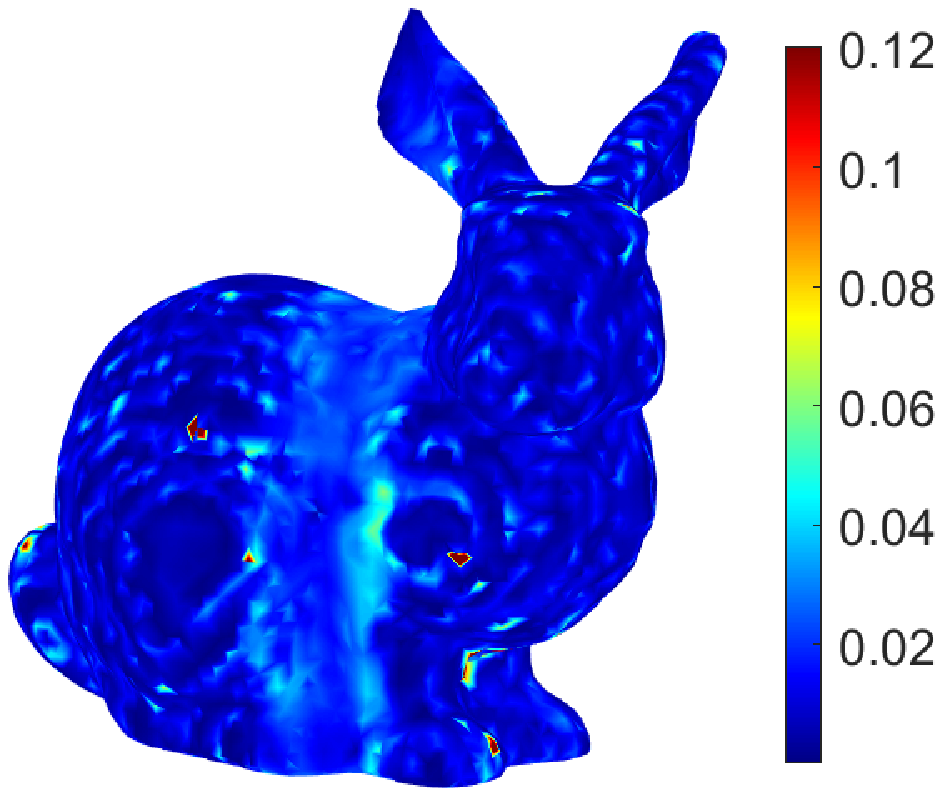}} \quad
\subfloat[$t\!=\!30 s$ \label{fig:multiscale_t30}]
{\includegraphics[trim=4cm 1.3cm 0.9cm 1.6cm,clip=true,width=0.35\linewidth]{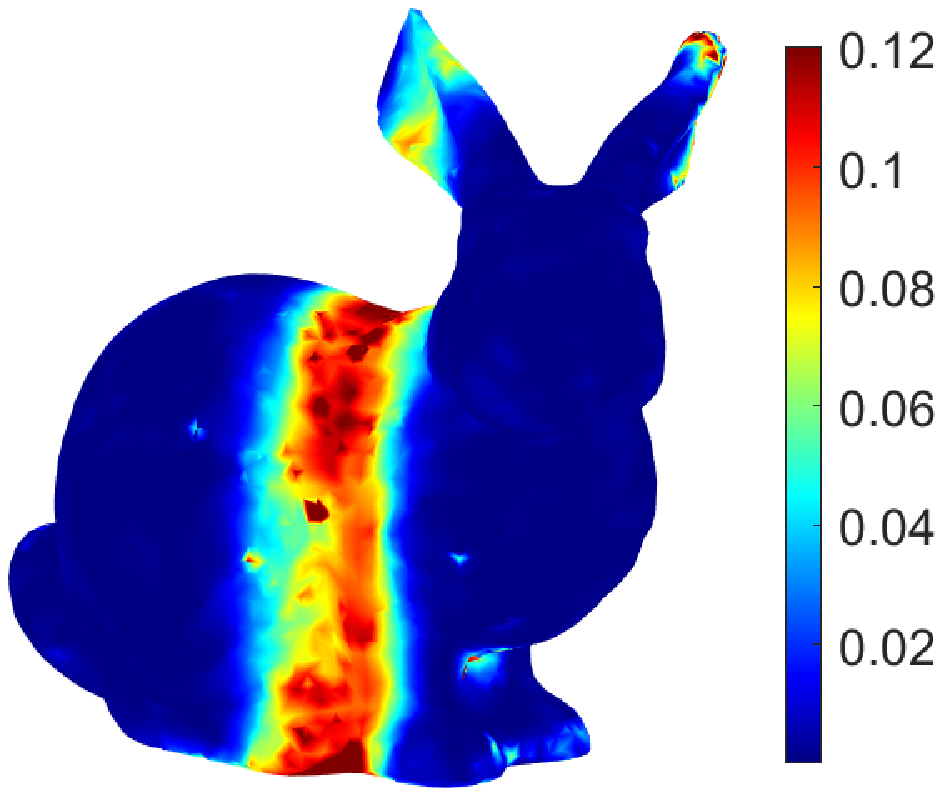}} \caption{Multiscale Analysis}
\vspace*{-0.6em}
\label{fig:multiscale_analysis}
\end{figure}

Firstly, we account a simple two levels signal defined on a triangular mesh representing the Stanford Bunny \cite{Turk} (see Fig \ref{fig:signal_a}). At this stage we have used $t\!=\!5s$ and $k\!=\!1$, Fig.\ref{fig:signal_b} shows the output $R^2[i]$, proving the effectiveness of the filters in capturing the variability of a simple signal. After that, we explain the multi-scale behavior by defining a test signal on the surface (see Fig.\ref{fig:multiscale_signal}) and filtering with two different $t$ values, $t_1 = 5s$ and $t_2 = 30s$. Taking into account the filter supports depicted in Fig.\ref{fig:heat_kernel}, it is easy to understand results reported in Fig.\ref{fig:multiscale_t5} and Fig.\ref{fig:multiscale_t30}: when a low-scale parameter $t$ is used, the filters are unable to detect spatially diffuse variability; conversely, wider support filters are suited to detect diffuse variations and not steep ones. This behavior paves the way towards multiscale pattern analysis.

\begin{figure}[t]
\centering
\subfloat[$n_x$]{
\includegraphics[trim=4cm 1.3cm 0.9cm 1.6cm,clip=true,width=0.28\linewidth]{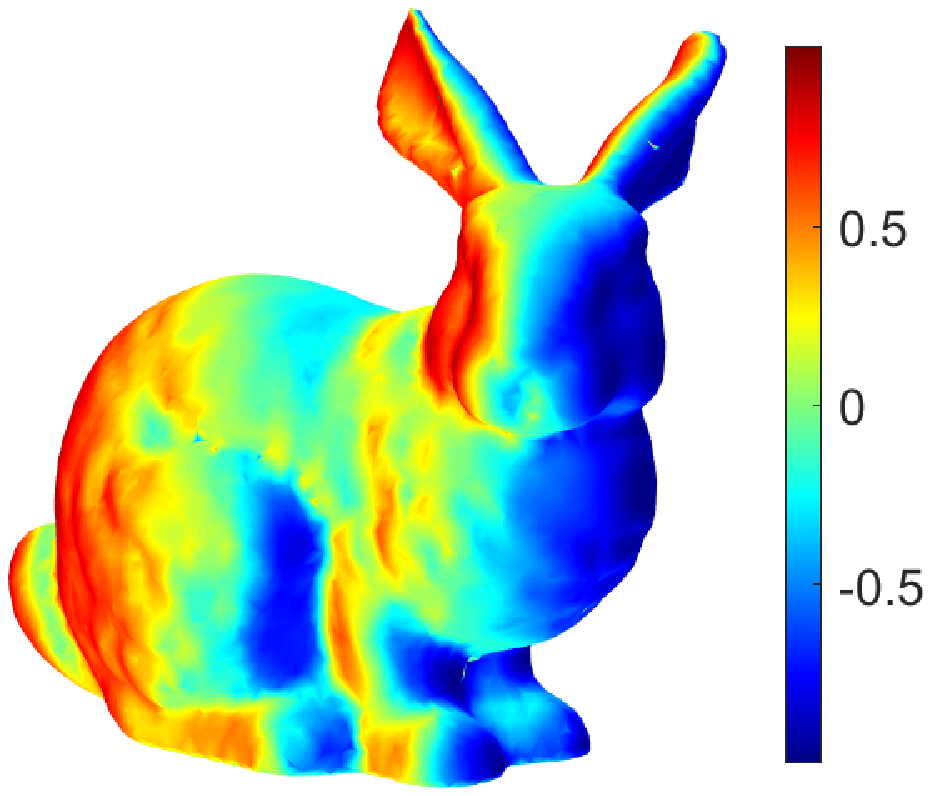}} \quad
\subfloat[$n_y$]{
\includegraphics[trim=4cm 1.3cm 0.9cm 1.6cm,clip=true,width=0.28\linewidth]{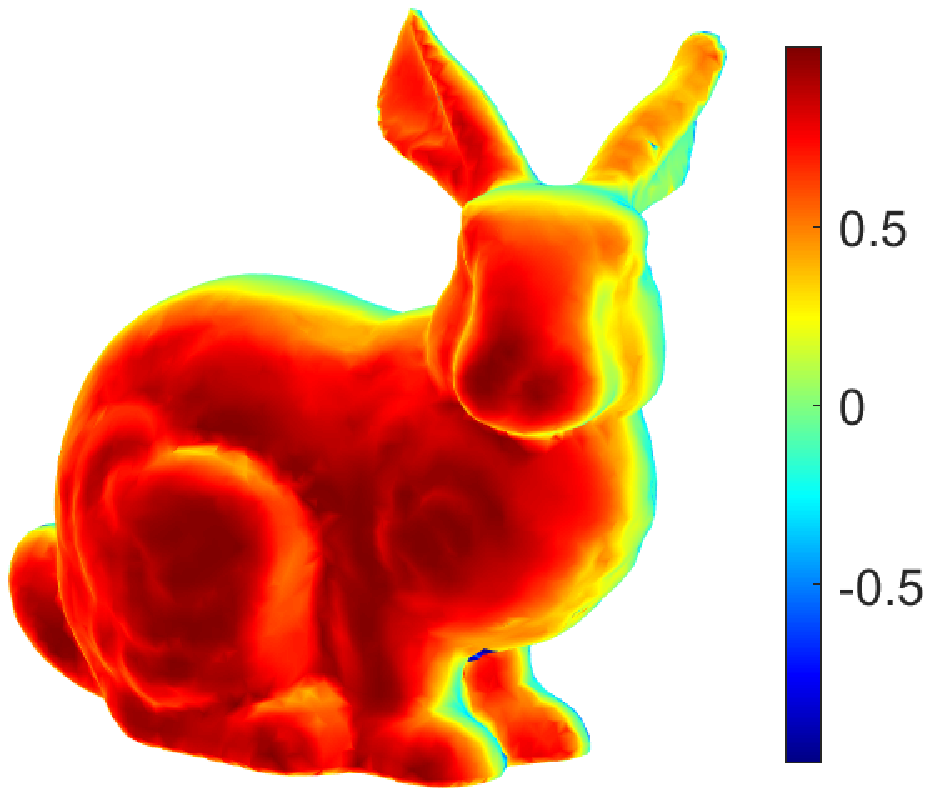}}
\subfloat[$n_z$]{
\includegraphics[trim=4cm 1.3cm 0.9cm 1.6cm,clip=true,width=0.28\linewidth]{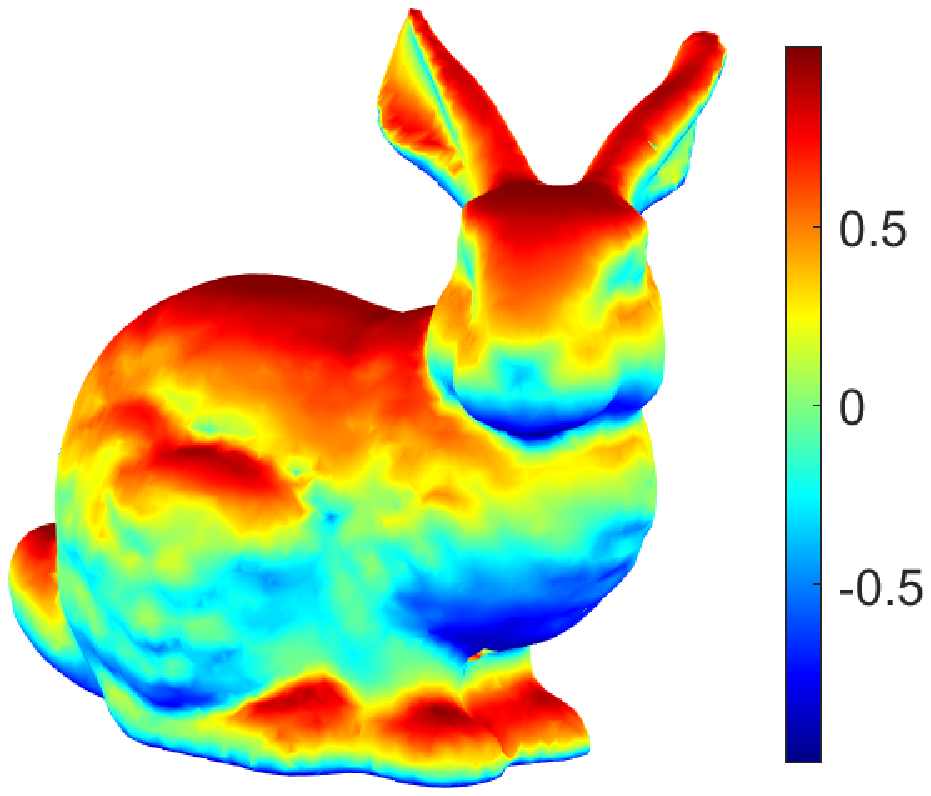}} \\
\subfloat[ ${R_x^2[i]}$]{
\includegraphics[trim=4cm 1.3cm 0.9cm 1.6cm,clip=true,width=0.28\linewidth]{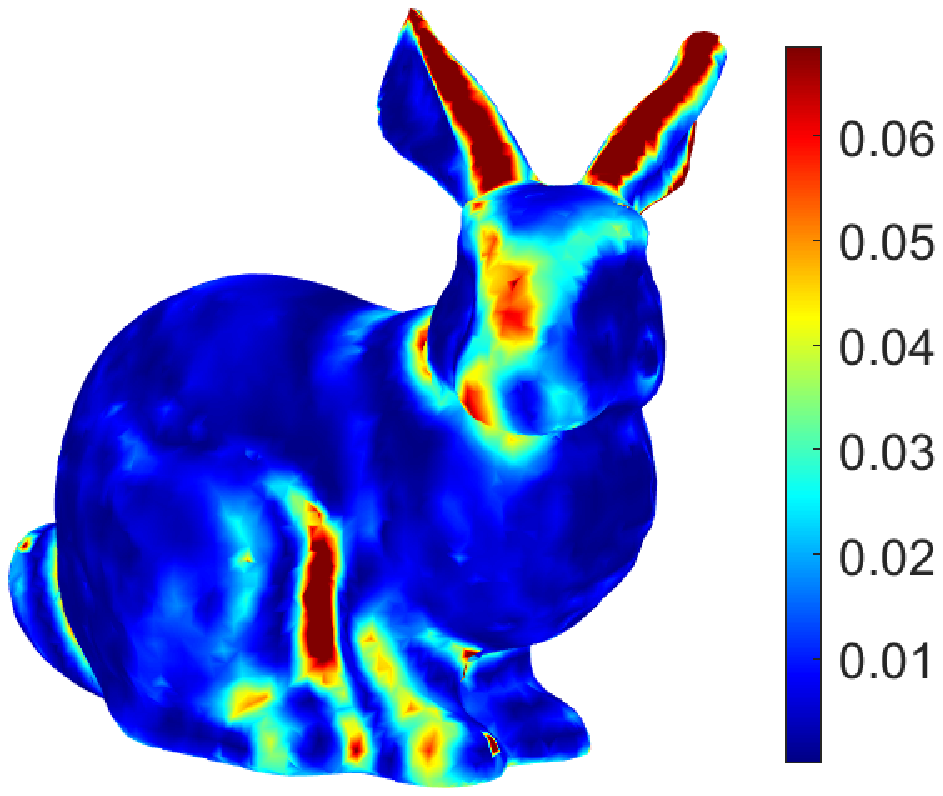}} \quad
\subfloat[${R_y^2[i]}$]{
\includegraphics[trim=4cm 1.3cm  0.9cm 1.6cm,clip=true,width=0.28\linewidth]{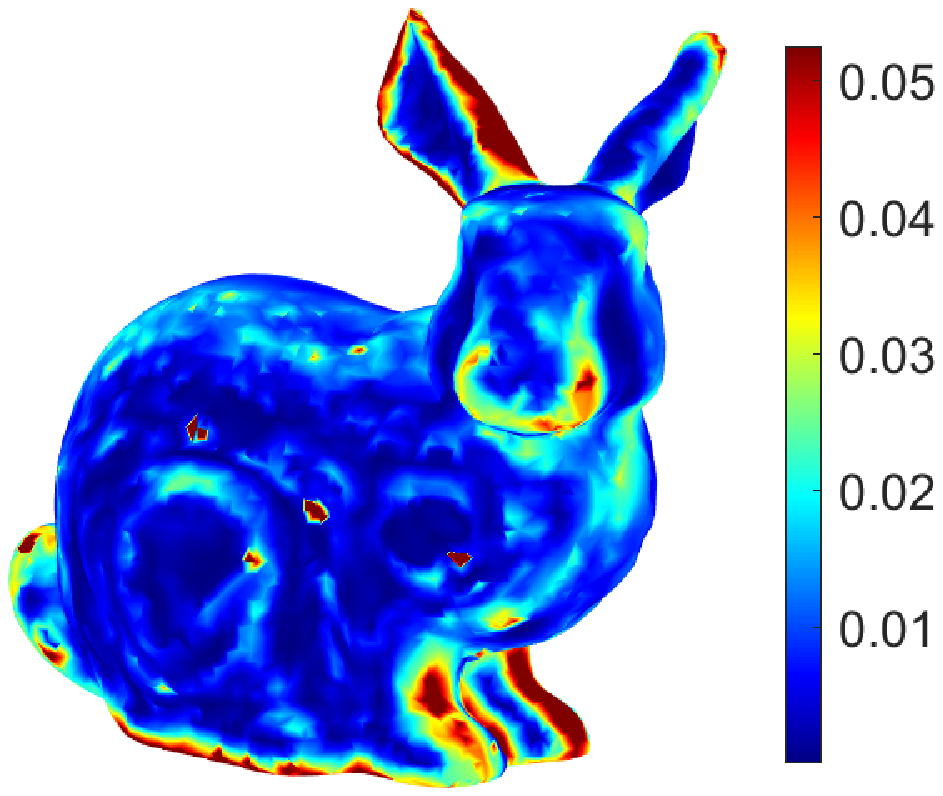}} \quad
\subfloat[${R_z^2[i]}$]{
\includegraphics[trim=4cm 1.3cm 0.9cm 1.6cm,clip=true,width=0.28\linewidth]{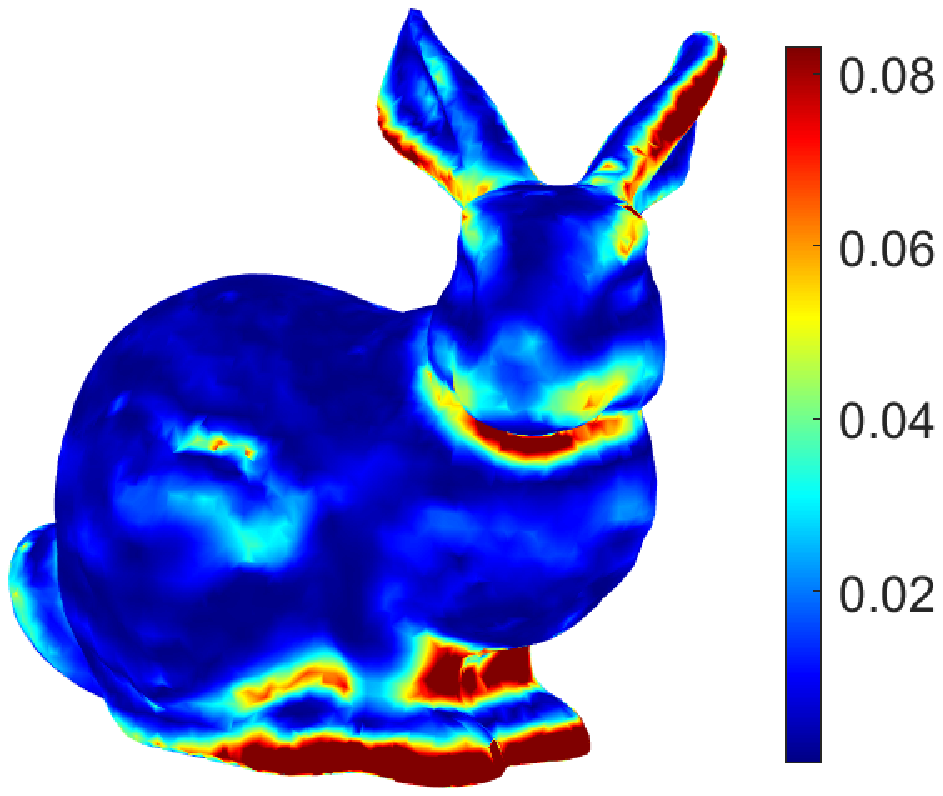}} 
\caption{MAHF applied to the normal vector's components is able to spot local curvature variation: $(a),(b),(c)$ normal components; $(d),(e),(f)$ filter's outputs.}
\vspace*{-1em}
\label{fig:normal}
\end{figure}

\begin{figure}[t]
\centering
\subfloat[MAHF ${R_N^2[i]}$\label{fig:our}]{
\includegraphics[trim=4cm 1.6cm 0.9cm 1.2cm,clip=true,width=0.4\linewidth]{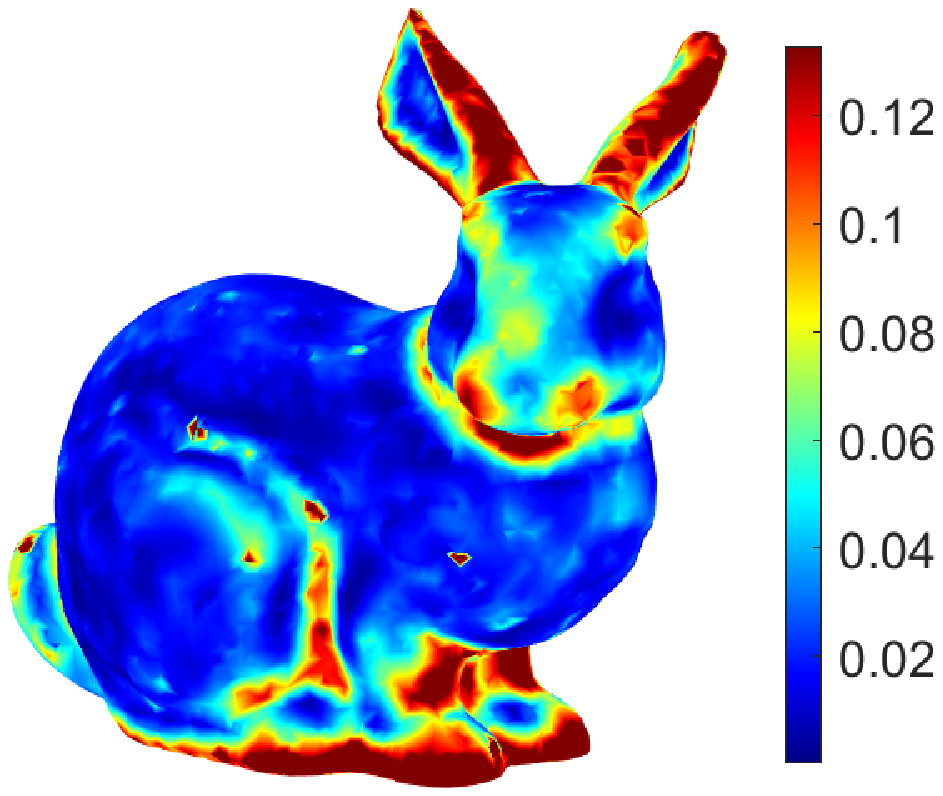}}
\subfloat[ MHW \cite{Kirgo}\label{fig:MHW}]{
\includegraphics[trim=4cm 1.6cm 0.9cm 1.2cm,clip=true,width=0.4\linewidth]{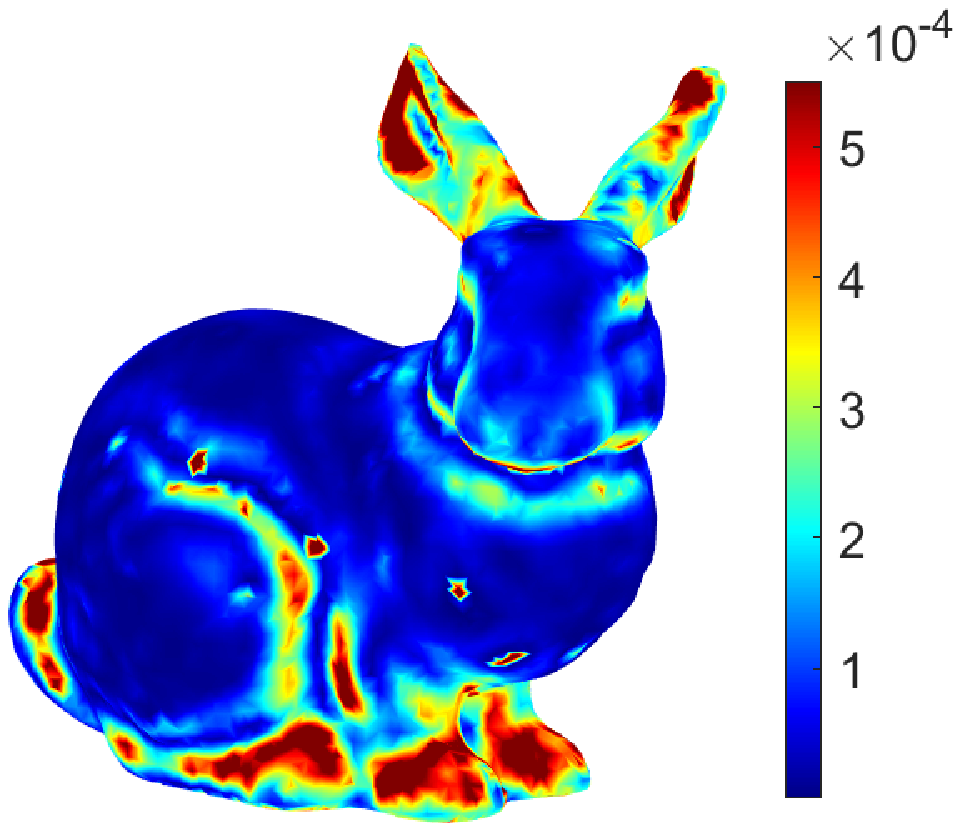}}
\caption{Normal field variation - Comparison}
\vspace*{-1em}
\label{fig:square}
\end{figure}

As a second demonstration, we aim to spot surface's local curvature changes, therefore we filter separately the normal components $\textbf{n}_p = [n_p^{(x)}, n_p^{(y)}, n_p^{(z)}]$ in a small scale fashion ($t\!=\!10s$). Fig.\ref{fig:normal} provides respectively the input signals and the corresponding results $[R_x^2[i], R_y^2[i], R_z^2[i]]$, confirming again the capability in variation extraction. Thus, in Fig.\ref{fig:our} we aggregate the separated normal components' variation as $R_N^2[i]\!=\!R_x^2[i] \!+\! R_y^2[i]\! +\! R_z^2[i]$, in order to spot global curvature changes. To further assess the performance, we repeated the filtering operation of normal's components using MHWs \cite{Kirgo, Tingbo} with the same scale factor ($t\!=\!10 s$), hence we aggregate again the results summing the squares of the filtered components. Fig.\ref{fig:MHW} shows the comparison, pointing out that our method is more suitable for edges detection, as can be seen around the neck, on the snout and in the bottom part of the legs, on which the normal field exhibits a large variation, correctly detected by our approach (Fig.\ref{fig:our}) but not by the MHWs (Fig.\ref{fig:MHW}).

\begin{figure}[t]
\centering
\subfloat[Face Mesh \label{fig:face_mesh}]{
\includegraphics[trim=3.5cm 1cm 0.7cm 0.7cm,clip=true,width=0.35\linewidth]{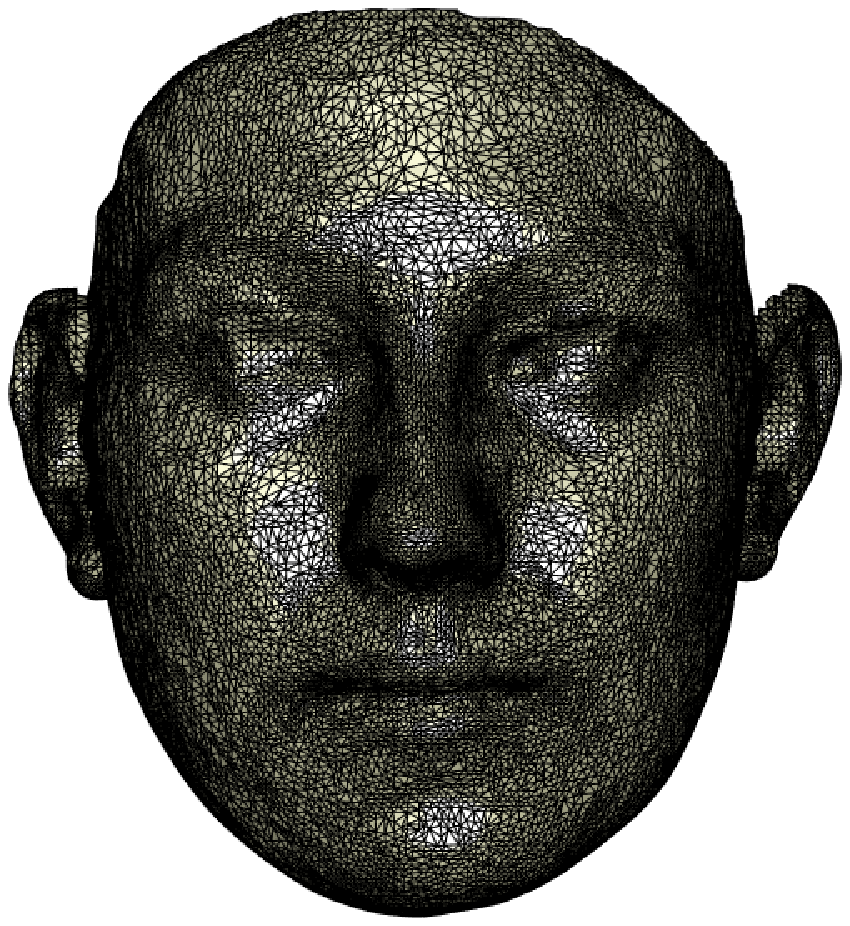}} \quad
\subfloat[Normal field variation \label{fig:face_norm}]{
\includegraphics[trim=3.5cm 1cm 1cm 0.7cm, width=0.35\linewidth]{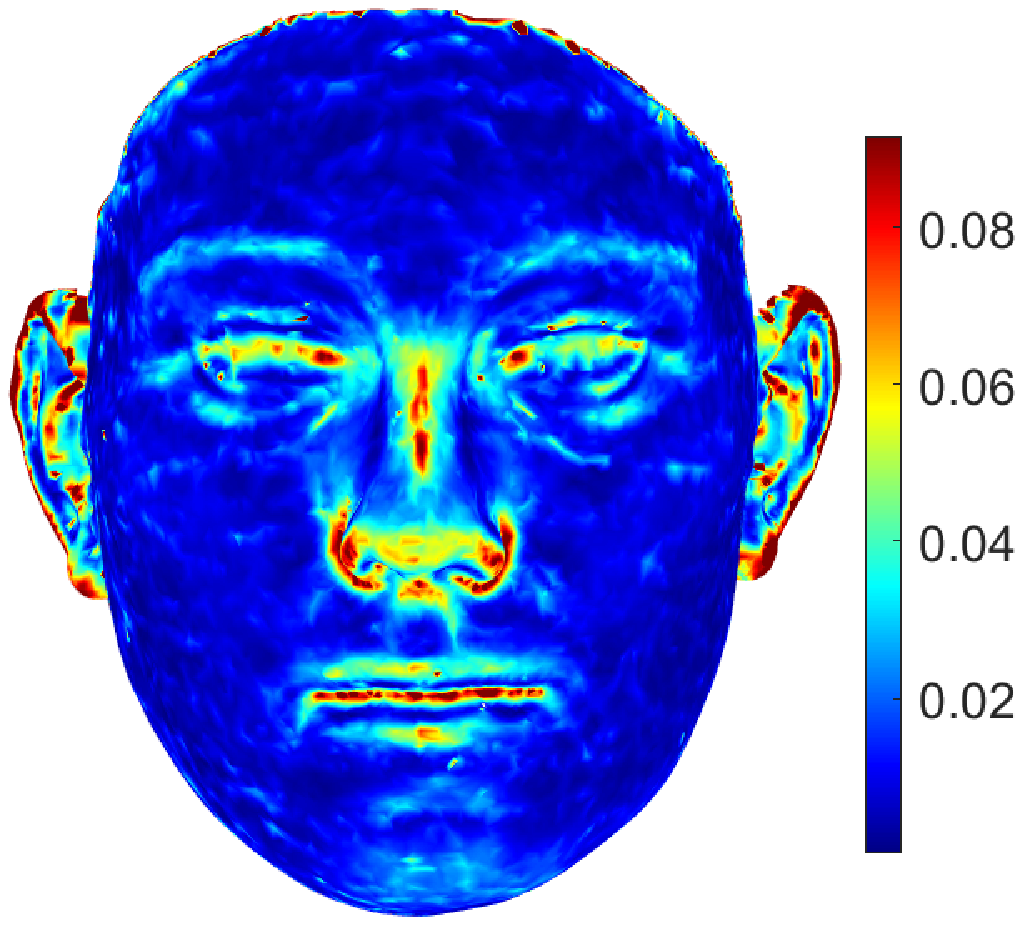}} 
\caption{MAHF applied to the normal vector's components is able to spot local curvature variation}
\vspace*{-1em}
\label{fig:face}
\end{figure}

\begin{figure}[t]
\centering
\subfloat[Textured Mesh\label{fig:face_text} ]{
\includegraphics[trim=3.5cm 1cm 1cm 0.7cm,clip=true,width=0.35\linewidth]{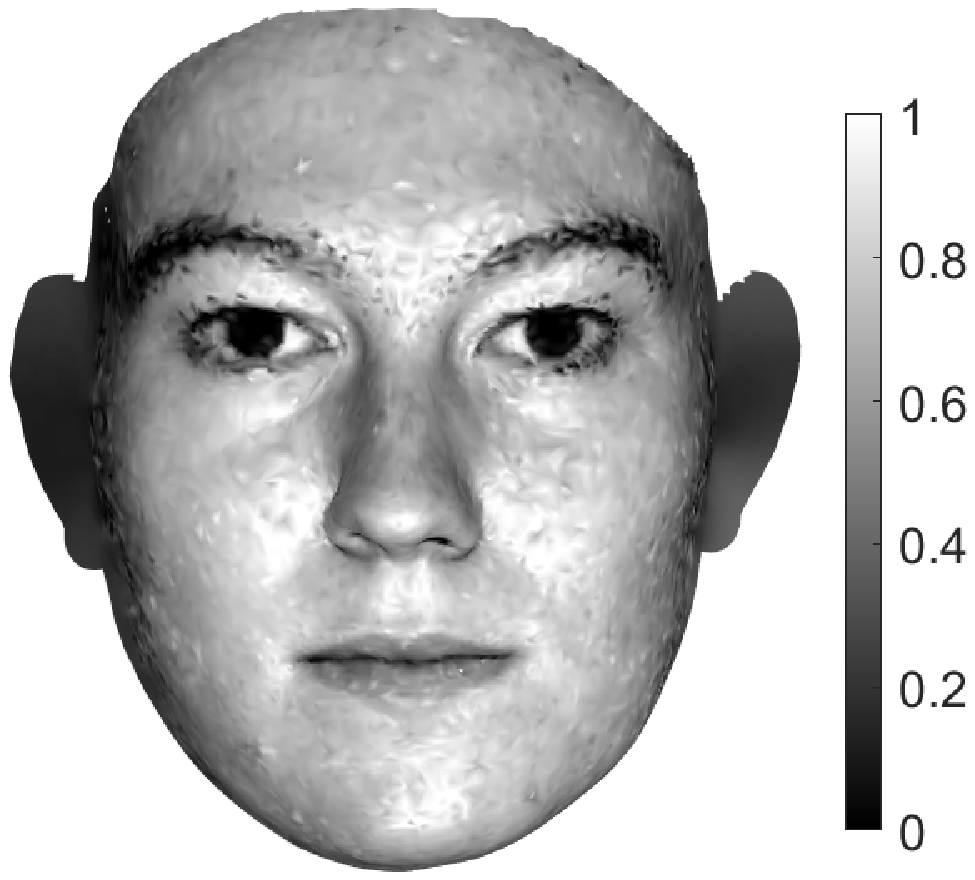}} \quad
\subfloat[Filtered Luminance\label{fig:face_lum_filt}]{
\includegraphics[trim=3.5cm  1cm 1cm 0.7cm,clip=true,width=0.35\linewidth]{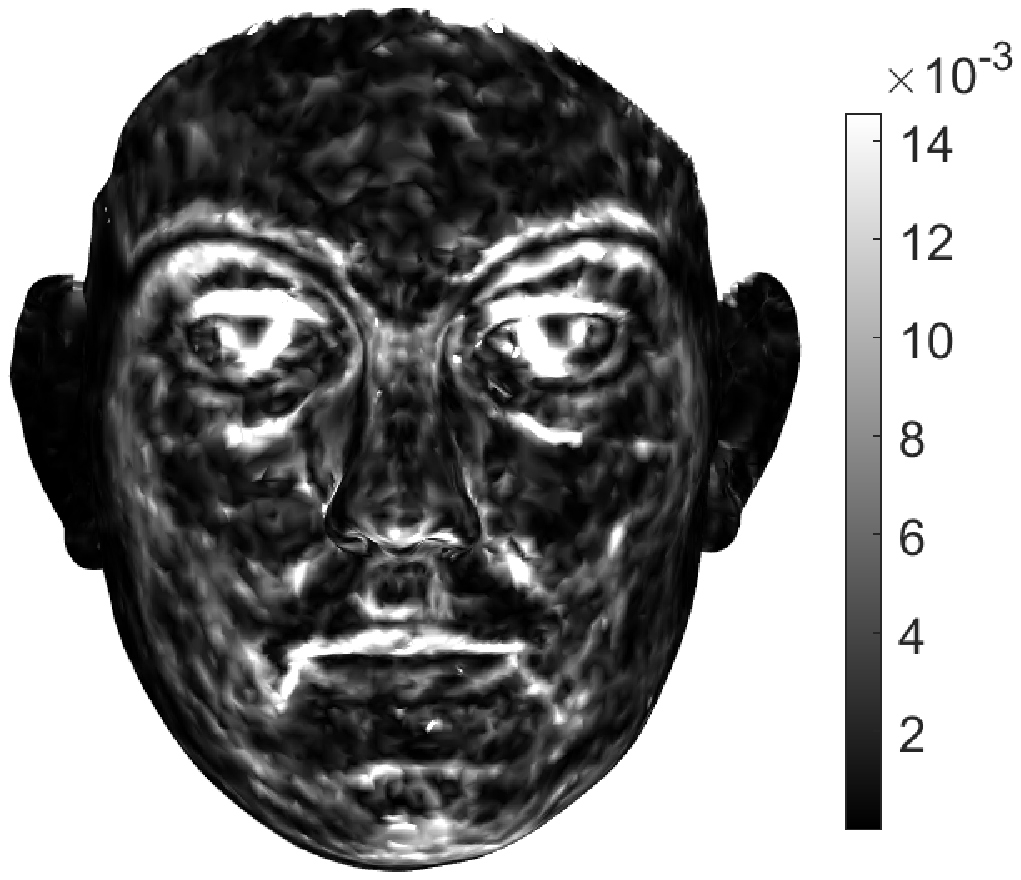}} 
\caption{MAHF applied to a luminance image textured on a face is able to spot local variation aware about mesh's support}
\label{fig:face_lum}
\end{figure}

\begin{figure}[h!]
\centering
\includegraphics[trim=3.5cm 1.5cm 0.9cm 0.9cm,clip=true,width=0.38\linewidth]{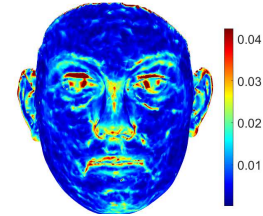}
\caption{Normal field and Luminance variation fusion, $\beta = \frac{1}{3}$}
\vspace*{-1em}
\label{fig:fusion}
\end{figure}

As a final demonstration, we apply MAHFs to signals defined over faces (presented FIDENTIS 3D Face Database \cite{fidentis}) represented as a meshed surface in Fig.\ref{fig:face_mesh}. Firstly, we repeat the procedure to evaluate the normal field variation $R_N^2[i]$, results are presented in Fig.\ref{fig:face_norm}. Then we move to the definition of a signal obtained texturing an image on the mesh vertices, shown in Fig.\ref{fig:face_text}. Filters' square module output $R^2_L[i]$ is presented in Fig.\ref{fig:face_lum_filt}:
we emphasize how luminance variations, such as those present on the pupils and lips, are highlighted, while nose and ears not, since the luminance has a constant value on these areas, on which variations are due to normal field rather than colours, and therefore they are correctly highlighted in \ref{fig:face_norm}. We emphasize how extracting information in a separate fashion can strengthen future applications with respect to different degradation on position or luminance. Hence, some applications could benefit from a fusion $R^2_L[i]+\beta R^2_N[i]$  between the two results, presented in Fig.\ref{fig:fusion}.

\section{Conclusion and Future Works}
\label{sec:concl}
In this work we have introduced \MAHF \:(MHF) for analysis of signals defined on  manifolds and on their discrete representations. MAHFs leverage heat diffusion equations and angular  metrics recently appeared in  different scientific communities. We have applied MAHF to extract visually relevant information on volumetric data in a multiple scale fashion. Future work is needed to gain  insight  about MAHFs' spectral properties, and their application in pre- post-processing stages. 

\bibliographystyle{IEEEtran}
\bibliography{IEEEabrv, mybib}

\end{document}